\documentclass[fleqn,twoside]{article}
\usepackage{espcrc2}

\usepackage{graphicx}
\usepackage[figuresright]{rotating}

\newcommand{\AmS}{{\protect\the\textfont2
  A\kern-.1667em\lower.5ex\hbox{M}\kern-.125emS}}

\def\lsim{\raise0.3ex\hbox{$\;<$\kern-0.75em\raise-1.1ex
\hbox{$\sim\;$}}}
\def\gsim{\raise0.3ex\hbox{$\;>$\kern-0.75em\raise-1.1ex
\hbox{$\sim\;$}}}

\title{
Establishing neutrino mass hierarchy and CP violation by two
identical detectors with different baselines using the J-PARC $\nu$ beam\thanks{Talk given by H. Nunokawa at NuFact05 workshop, 21-26 June 2005, Frascati, Italy, based on \cite{twinHK}.}
}

\author{Masaki Ishitsuka\address{Research Center for Cosmic Neutrinos, 
Institute for Cosmic Ray Research, University of Tokyo,  \\ 
Kashiwa, Chiba 277-8582, Japan}%
        \thanks{Present address: Indiana University, Bloomington, IN, USA
},
        Takaaki Kajita\addressmark,
        Hisakazu Minakata\address{
Department of Physics, Tokyo Metropolitan University, 
Hachioji, Tokyo 192-0397, Japan }
        and
        Hiroshi Nunokawa\address{
Departamento de F\'{\i}sica, Pontif{\'\i}cia Universidade Cat{\'o}lica 
do Rio de Janeiro, \\
C. P. 38071, 22452-970, Rio de Janeiro, Brazil
}}
       
\begin{document}

\begin{abstract}
We discuss how and to what extent one can determine 
the neutrino mass hierarchy, 
normal or inverted, 
and at the same time uncover CP violation in the lepton sector 
by using two identical detectors with different baselines 
in neutrino oscillation experiments using low energy 
superbeam from the J-PARC facility. 

\vspace{1pc}
\end{abstract}

\maketitle

\section{Introduction}
In the neutrino sector, we still do not know 
the value of $\theta_{13}$, the sign of $\Delta m^2_{31}$, 
which is positive (negative) if the mass hierarchy 
is normal (inverted), and the value of the CP violating 
phase $\delta$. 
It is well known that the determination of these parameters 
by accelerator-based neutrino oscillation experiments 
suffer from the ambiguities coming from so called the 
parameter degeneracy~\cite{othertalks}.
In this work, we discuss possible way of determining the type 
of the neutrino mass hierarchy as well as the CP violating phase 
simultaneously by resolving the degeneracy using two identical detectors 
with different baselines. 

\section{Principle of two-detector measurement}

We start from the original proposal of phase II of the J-PARC neutrino project~\cite{JPARC} in which an off axis neutrino beam with 4 MW beam 
power and a megaton (Mton) water Cherenkov detector, Hyper-Kamiokande (HK),
whose fiducial volume is 0.54 Mton will be used to measure 
appearance events in 
$\nu_{\mu} \rightarrow \nu_{\rm e}$ and 
$\bar{\nu}_{\mu} \rightarrow \bar{\nu}_{\rm e}$ channels. 
We then propose a ``minor'' modification; 
Instead of placing a 1 Mton HK at Kamioka with 
the baseline $L=295$ km, we propose 
to divide the detector into two identical half Mton detectors 
with fiducial volume 0.27 Mton, and 
place one of them in Kamioka and the other somewhere in Korea 
with the baseline $L=1050$ km.

Assuming the same performance for the two identical detectors, 
most of the systematic errors cancel between the detectors.  
Furthermore, we expect that the neutrino energy spectra are the 
same at the two sites without oscillation because of the same off axis 
angle of 2.5$^\circ$ both at Kamioka and Korean detectors with 
peak energy of 650 MeV.  
It is possible under the current design of the J-PARAC neutrino beam line. 
The overall normalization simply scales as 
$1/L^2 \sim 1/10$.

Such an experimental set up implies that one can perform clean 
detection of the distortion of neutrino energy spectrum caused 
purely by the oscillation effect, if the difference in background due to 
different conditions at the two sites is tolerable. 
Namely, any differences of the $\nu$ energy spectra among the front 
and the two half Mton 
detectors are due to vacuum oscillation which is sensitive to $\delta$ 
at such low energy, and to the matter effect which is crucial to 
determine the hierarchy, the sign of $\Delta m^2_{31}$.

\section{Analysis Results}

\begin{figure}[htb]
\vglue -0.5cm
\includegraphics[width=0.44\textwidth]{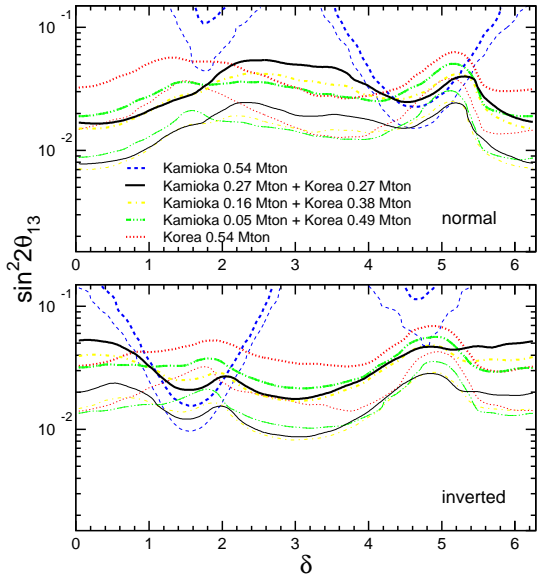}
\vglue -1.0cm
\caption{ 2(thin lines) and 3(thick lines) 
standard deviation sensitivities to  the mass hierarchy for 
the various fiducial volume ratio of 
[Kamioka: Korea] = [1:0] (dashed lines, blue), [1:1] (solid lines, black), 
[3:7] (dash-dot lines, yellow), [1:9] (dash-dot-dot lines, green), 
and [0:1] (dotted lines, red), keeping the total volume equals to 0.54 Mton.
4 years running with neutrino beam and another 4 years with 
anti-neutrino beam are assumed.
The other mixing parameters are fixed to the current best fit 
values as described in \cite{twinHK}. 
}
\vglue -0.5cm
\label{fig:sensitivity_mass}
\end{figure}

In order to quantify the determination power of 
the mass hierarchy as well as the CP violating phase, 
we performed a detailed $\chi^2$ analysis. 
We present in Fig.\ref{fig:sensitivity_mass} the parameter region 
in the $\delta-\sin^2 2\theta_{13}$ plane in which the 
mass hierarchy can be determined at 2$\sigma$  and 3$\sigma$ CL. 
One can establish the type of the mass hierarchy in any region 
above these curves.  
In fact, we have also examined the cases of various volume ratios 
of the two detectors keeping the total volume equals to 0.54 Mton, 
as reported in Fig.\ref{fig:sensitivity_mass}.

\begin{figure}[htb]
\vglue -0.5cm
\includegraphics[width=0.44\textwidth]{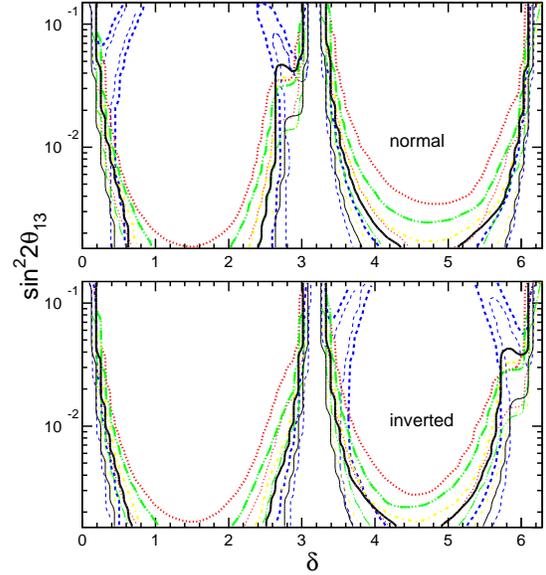}
\vglue -1.0cm
\caption{ Same as Fig.\ref{fig:sensitivity_mass} 
but for the CP violation (see the text). 
}
\vglue -0.9cm
\label{fig:sensitivity_cp}
\end{figure}

In Fig.\ref{fig:sensitivity_cp} we show the similar curves but for the CP violation. 
We can establish the CP violation (if $\delta$ is
different from 0 or $\pi$) in any region above these curves.  
From Figs.\ref{fig:sensitivity_mass} and \ref{fig:sensitivity_cp}
we conclude that the option of the 
two detectors with fiducial volume of 0.27 Mton each 
at Kamioka and Korea seem to be close to the optimal. 

\section{Conclusions}
We have demonstrated that the two-detector complex can determine 
neutrino mass hierarchy down to 
$\sin^2 2\theta_{13} \gsim 0.03$ (0.055) for any value of $\delta$ 
at 2$\sigma$ (3$\sigma$) CL, as indicated in Fig.\ref{fig:sensitivity_mass}.
It should be noted that the sensitivity to the CP violation of 
the current design 
of J-PARC phase II project is essentially kept or even enhanced 
at  $\sin^2 2\theta_{13} \gsim 0.01$. 
See Ref.~\cite{twinHK} for more details of this work.

\end{document}